\documentclass[sn-mathphys,Numbered,iicol]{sn-jnl}% Math and Physical Sciences Reference Style
\usepackage{graphicx}%
\usepackage{multirow}%
\usepackage{amsmath,amssymb,amsfonts}%
\usepackage{amsthm}%
\usepackage{mathrsfs}%
\usepackage[title]{appendix}%
\usepackage{xcolor}%
\usepackage{textcomp}%
\usepackage{manyfoot}%
\usepackage{booktabs}%
\usepackage{algorithm}%
\usepackage{algorithmicx}%
\usepackage{algpseudocode}%
\usepackage{listings}%
\usepackage{gensymb}%%%%

\theoremstyle{thmstyleone}%
\theoremstyle{thmstyletwo}%

\theoremstyle{thmstylethree}%

\begin{document}

\title[Article Title]{Fast and Efficient Type-II Phototransistors Integrated on Silicon}

\author[1]{\fnm{Lining} \sur{Liu}}
\author[1]{\fnm{Simone} \sur{Bianconi}}
\author[1]{\fnm{Skylar} \sur{Wheaton}}
\author[1]{\fnm{Nathaniel} \sur{Coirier}}
\author[2]{\fnm{Farah} \sur{Fahim}}
\author[1]{\fnm{Hooman} \sur{Mohseni}}\email{hmohseni@northwestern.edu}

\affil[1]{\orgdiv{Bio-Inspired Sensors and Optoelectronics Laboratory}, \orgname{Northwestern University}, \orgaddress{\street{2145 Sheridan Rd}, \city{Evanston}, \postcode{60208}, \state{Illinois}, \country{USA}}}
\affil[2]{\orgdiv{ASIC Development Group, Particle Physics Division}, \orgname{Fermi National Accelerator}, \orgaddress{\city{Batavia}, \postcode{60510}, \state{Illinois}, \country{USA}}}


\abstract{Increasing the efficiency and reducing the footprint of on-chip photodetectors enables dense optical interconnects for emerging computational and sensing applications. Avalanche photodetectors (APD) are currently the dominating on-chip photodetectors. However, the physics of avalanche multiplication leads to low energy efficiencies and prevents device operation at a high gain, due to a high excess noise, resulting in the need for electrical amplifiers. These properties significantly increase power consumption and footprint of current optical receivers. In contrast, heterojunction phototransistors (HPT) exhibit high efficiency and very small excess noise at high gain. However, HPT’s gain-bandwidth product (GBP) is currently inferior to that of APDs at low optical powers. Here, we demonstrate that the type-II energy band alignment in an antimony-based HPT results in a significantly smaller junction capacitance and higher GBP at low optical powers. We used a CMOS-compatible heterogeneous integration method to create compact optical receivers on silicon with an energy efficiency that is about one order of magnitude higher than that of the best reported integrated APDs on silicon at a similar GBP of $\sim$270 GHz. Bitrate measurements show data rate spatial density above 800 Tbps/mm2, and an energy-per-bit consumption of only 6 fJ/bit at 3 Gbps. These unique features suggest new opportunities for creating highly efficient and compact on-chip optical receivers based on devices with type-II band alignment.}

%%================================%%
%% Sample for structured abstract %%
%%================================%%




\maketitle

\section{Introduction}\label{sec1}

Integrated photodetectors have attracted great interest for a wide variety of applications, including photonic neural network accelerators \cite{1_pai2023experimentally,2_feldmann2021parallel}, photonic signal processors\cite{3_liu2016fully}, 3D depth imaging \cite{4_lee2023caspi}, and optical interconnection \cite{5_michel2010high,6_shin2015control,7_miller1997physical}. Particularly, optical interconnects are considered to be one of the most promising candidates for next-generation on-chip communication \cite{7_miller1997physical} due to their unique advantages \cite{8_han2017efficient,9_virot2014germanium,10_crosnier2017hybrid}. However, current on-chip implementations fail to meet the footprint and energy consumption requirements necessary to surpass the performance of electrical interconnects \cite{7_miller1997physical,11_miller2009device}. Energy efficiency and energy-per-bit are key figures of merit that limit the growth in data storage, transmission, and computation, hence improving both these metrics in key components is essential \cite{11_miller2009device,12_miller2017attojoule,13_rumley2015silicon,14_bahadori2017energy}. In this work we focus on the receiving end of the on-chip data transmission, and present a new high-performance integrated optical receiver with significant improvements in both energy efficiency and energy-per-bit over the existing optical receivers integrated on silicon.
In order to drive meaningful on-chip loads with the minimum input optical signal, an integrated receiver requires some form of amplification of the detected signal (see Supplementary, Figure S6). Detectors without internal gain mechanisms (e.g. p-i-n photodiodes) rely entirely on external amplifiers; however, even the latest generations of custom-designed high-speed CMOS amplifiers \cite{15_gasser2023highly} require significant power and on-chip area (See Supplementary, Table SIII). Conversely, most state-of-the-art optical receivers such as APD and HPT utilize internal amplification to boost the detected signal in order to either directly drive a load, or to reduce the complexity of the external amplifiers \cite{16_huang201625,17_chen2015high}. In this configuration, higher internal gains enable to either directly drive larger loads, or to eliminate the need for external amplifiers. The large excess noise factor of APDs limits their capability in operating at high gain, hence external amplifiers are typically still necessary; conversely, HPTs can achieve large internal amplification at high speed with relatively low excess noise and dark current. Notably, staircase APDs have achieved a low excess noise factor \cite{18_march2021multistep} but the maximum achievable gain has remained small since the invention of these devices about four decades ago \cite{19_williams1982graded}.
Early studies suggested that HPTs could be used for high-speed receivers \cite{20_chandrasekhar1993high}, and by the year 2000 they achieved bitrates of 40 Gbps \cite{21_huber2000inp}. Unfortunately, these high-speed HPT receivers suffered from low sensitivity (needed about -8 dBm optical power), and they were not integrated on silicon. These issues have moved the researchers’ focus towards APD receivers in the past decade \cite{9_virot2014germanium,22_chen2015high,23_benedikovic202040,24_zeng2019silicon}. However, beside the limitations on gain discussed above, most CMOS-compatible APDs also require a large bias voltage to achieve proper gain values, and exhibit high dark current \cite{9_virot2014germanium,17_chen2015high,25_assefa2010reinventing,26_kang2009monolithic,27_kang2011high}, leading to poor energy efficiencies and large power consumption. 
Here, we demonstrate that using a type-II energy band alignment in a properly designed HPT leads to devices that can simultaneously achieve a high gain-bandwidth product (GBP) at low optical powers, a high sensitivity, and a high energy efficiency. Our type-II heterojunction phototransistor (T2-HPT) integrated on silicon achieve a GBP similar to the best reported APDs on silicon, but with much lower dark current and operating bias voltage than APDs, much smaller footprint that enables dense integration, and about ten times higher energy efficiency for a given load and similar maximum bitrate. More importantly, we show that it achieves such a high speed at a low optical power — well below that of the best HPTs previously reported \cite{28_thuret1999high,29_campbell1981small,30_fritzsche1981fast,31_chandrasekhar1991demonstration}. Our experimental and modeling results suggest that the superior performance of these devices is due to the lower capacitance of type-II heterojunctions compared to the traditional type-I heterojunctions, as well as to the better thermal management of our integration method.
All results presented here are based on devices integrated on silicon and optically coupled to on-chip hydrogenated amorphous silicon (a-Si:H) waveguides and grating couplers using a CMOS-compatible process (see Methods). Since the integration approach is not limited to a specific CMOS process, and the optical interconnects (i.e. waveguides and couplers) are not substrate-specific, our approach can be used with most existing CMOS technologies. 

\begin{figure*}[t]
    \centering
    \includegraphics[width=12.5cm]{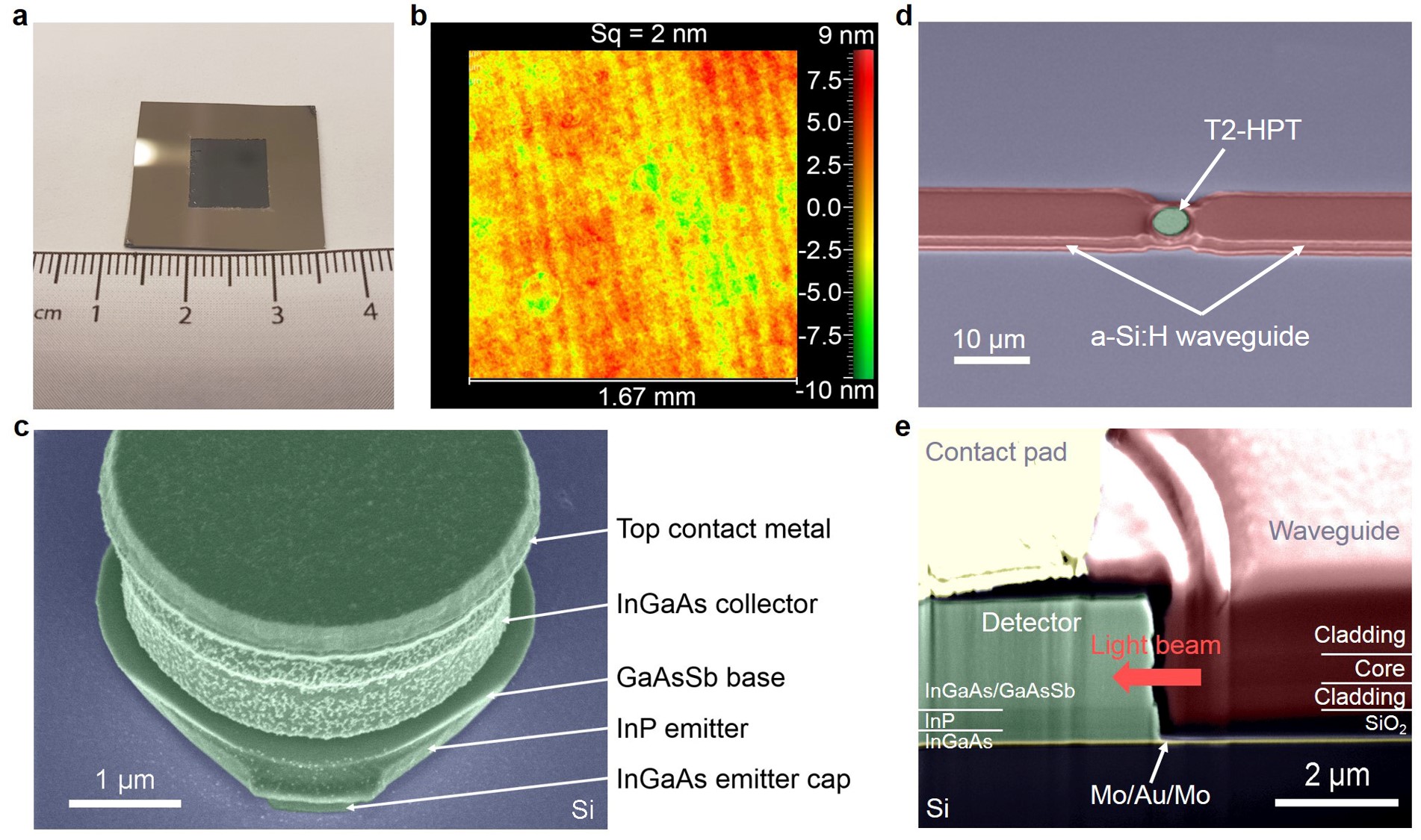}
    \caption{Images of the integrated T2-HPT. \textbf{a}, Photograph of the InP/GaAsSb/InGaAs epitaxial layer transferred to silicon, and after the InP substrate was removed. \textbf{b}, Surface morphology of the transferred material shows rms roughness of about 2 nm. \textbf{c}, SEM image of the detector after pillar etching shows the epitaxial III-V semiconductor layers on the silicon substrate. \textbf{d}, SEM of the integrated T2-HPT after waveguide etching and before the BCB planarization. \textbf{e}, Cross-sectional FIB-SEM image of an integrated device at the end of the fabrication process.}
    \label{fig1}
\end{figure*} 

\section{Results}\label{sec2}

\subsection{Device Design}\label{subsec1}
We used a low-temperature Mo/Au-Au/Mo bonding method \cite{32_park2018ingaas} to transfer epitaxial layers to silicon substrates (See Supplementary I). We observed a typical root-mean-square (rms) roughness of $\sim$2 nm over large areas of the transferred film, indicating high-quality bonding (Fig. \ref{fig1}a-b). After the transfer, T2-HPTs with different sizes (ranging from 100 \textmu m to 2 \textmu m in diameter) were fabricated on the silicon substrate and integrated with a-Si:H waveguides and grating couplers (See Methods). a-Si:H can be easily deposited at low temperatures ($\leq$ 300 $\degree$C) in a process compatible with III-V materials and back end of line (BEOL) CMOS processes \cite{33_roy2019design,34_jiang2018cmos}. Our approach allows fine control of the a-Si:H refractive index by tuning the RF power and pressure during deposition \cite{35_kwakernaak2006amorphous} (see Methods and Supplementary I). Fig. \ref{fig1}c shows a scanning electron microscope (SEM) image of a T2-HPT device before the waveguide integration, with clear collector, base and emitter layers of the HPT. The SEM images of a detector after waveguide integration show a top (Fig. \ref{fig1}d) and a cross-sectional (Fig. \ref{fig1}e) view of the interface between the waveguide and the detectors.

\subsection{Device Measurement}
\begin{figure*}[t]
    \centering
    \includegraphics[width=13cm]{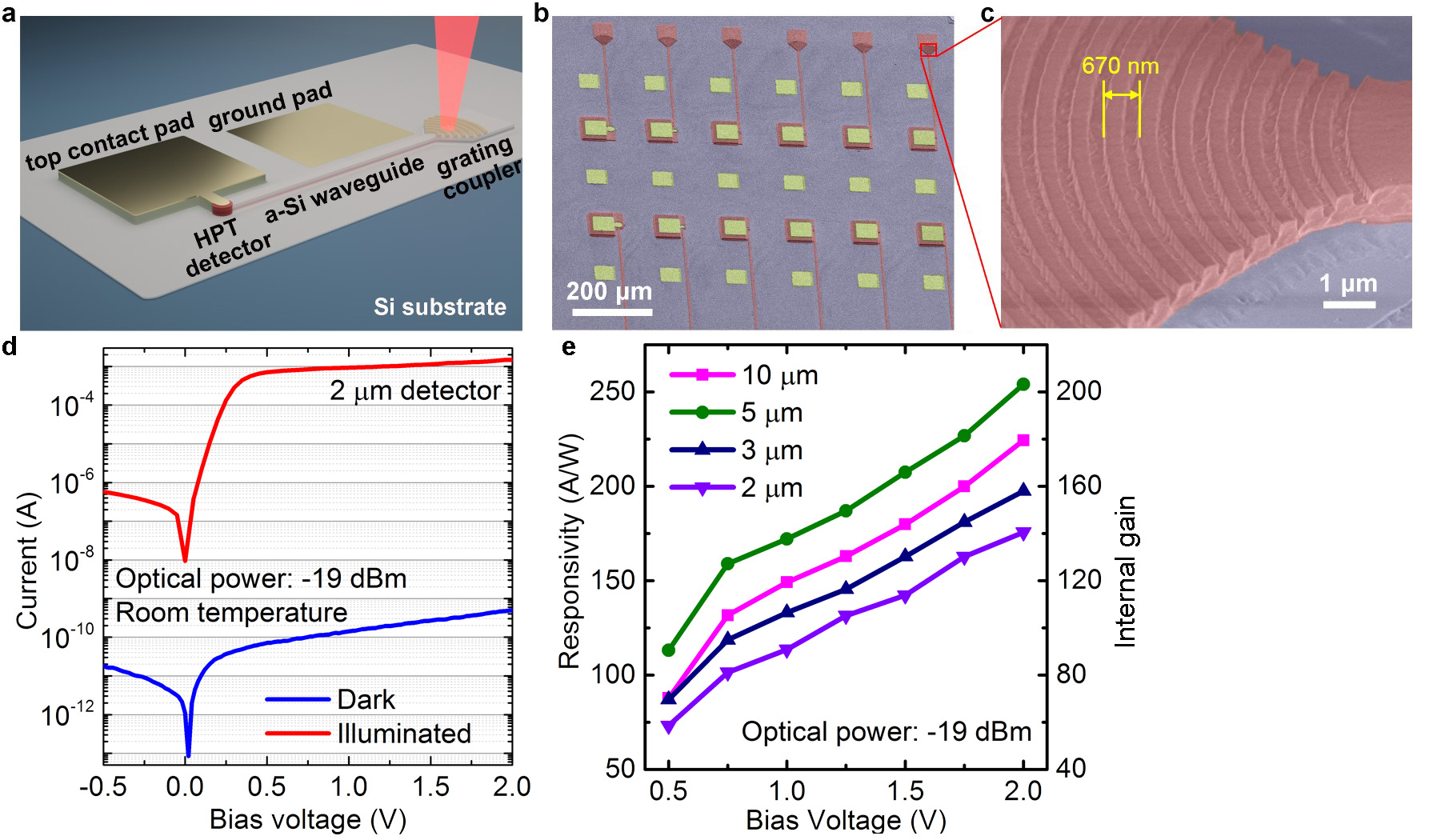}
    \caption{Integrated T2-HPT detectors with top coupling. \textbf{a}, Schematic diagram of the measurement setup with the grating coupler. \textbf{b}, SEM image of the integrated detectors of different sizes. \textbf{c}, SEM image of the zoomed-in grating coupler with a period of 670 nm. \textbf{d}, Current-voltage characteristic of the integrated 2 \textmu m detectors at room temperature at dark and under illumination. \textbf{e}, DC responsivity R of the detectors of different sizes as a function of bias voltage at an optical power of -19 dBm coupled to the detector. }
    \label{fig2}
\end{figure*}

\label{subsec2
}
We characterized the fully integrated T2-HPT devices with the architecture shown in Fig. \ref{fig2}a. While we measured devices with different diameters, here we mainly focus on the performance of devices of 2 \textmu m diameter. We measured the dark current ($I_d$) and the photocurrent ($I_{ph}$) of these devices at room temperature by coupling the free-space collimated output of a 1.55 \textmu m laser diode into the grating couplers (See Fig. \ref{fig2}b and Fig. \ref{fig2}c). Fig. \ref{fig2}d shows the dark and illuminated current of a 2 \textmu m T2-HPT. The device dark current at 2 V is about 0.5 nA, which is several orders of magnitude lower than the dark current of the best integrated APDs at their operating bias voltages. This is despite the dark current density increasing when shrinking the device diameters under 10 \textmu m, due to a significant contribution from surface effects, which suggests that dark currents could be reduced with better surface passivation (See supplementary II). The photocurrent shown in Fig. \ref{fig2}d was measured at an optical power of -19 dBm coupled to the 2 \textmu m T2-HPT. Fig. \ref{fig2}e shows the DC responsivity at this power, exceeding 100 A/W at voltages as low as 0.75 V. Devices smaller than 5 \textmu m have lower responsivity, most likely due to the reduced overlap with the optical mode of the 5 \textmu m-wide waveguides. This effect could be prevented by decreasing the width of the waveguides or introducing tapered waveguides. Furthermore, the low coupling to small devices can be addressed by employing hybrid optical antennae in order to enhance the local density of states and coupling of the near-field to the waveguide modes \cite{36_bonakdar2014impact,37_tang2008nanometre,38_melikyan2014high}.

\begin{figure*}[t]
    \centering
    \includegraphics[width=12.5cm]{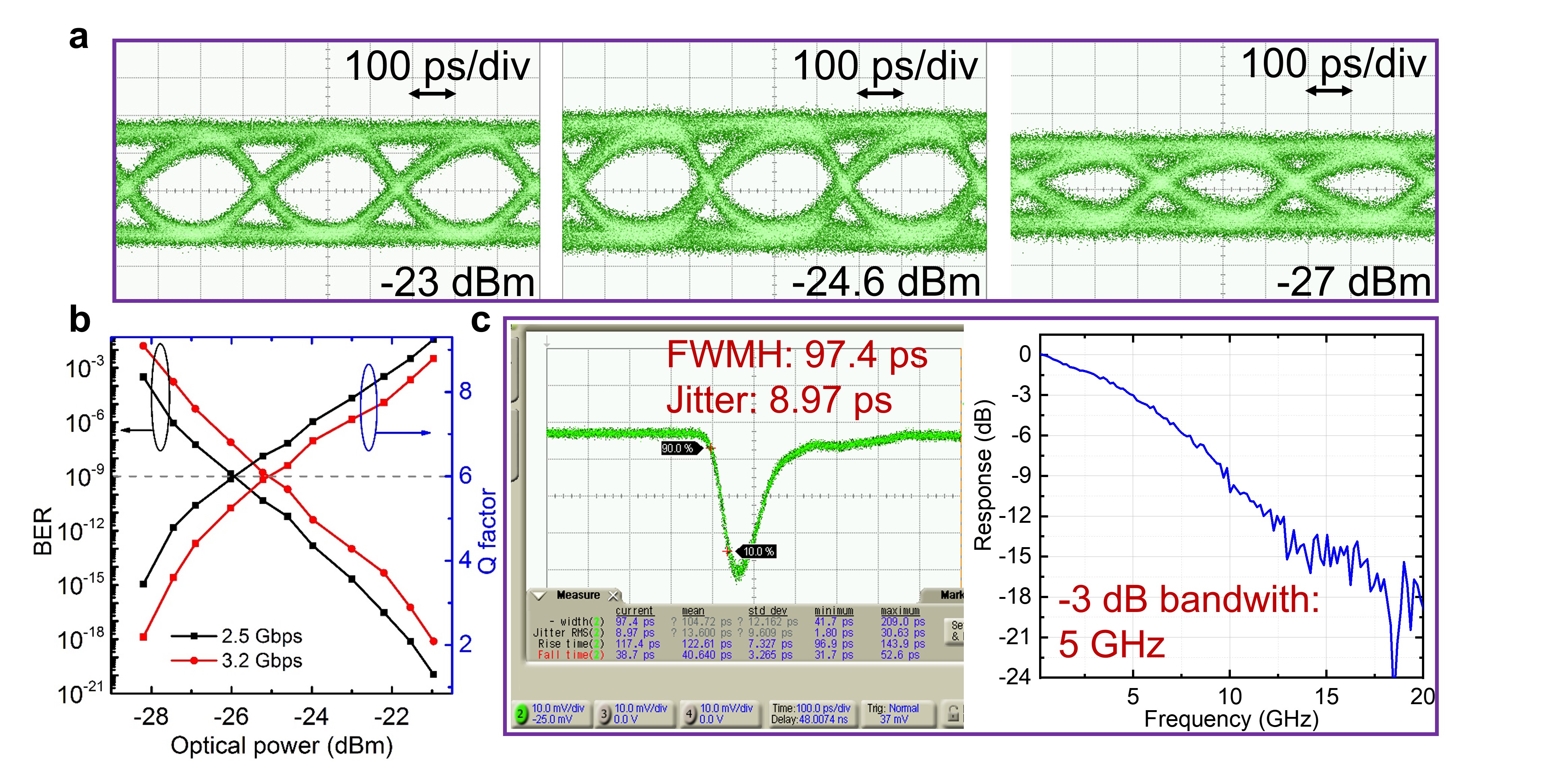}
    \caption{Speed measurement of the integrated T2-HPT detectors. \textbf{a}, Eye diagrams of 2 \textmu m device at 3.2 Gbps data rate, at different optical powers coupled to the detector. \textbf{b}, Bit error rate and Q factor as a function of optical power, extracted from the eye diagrams of the 2 \textmu m detector. \textbf{c}, Output pulse from the integrated 2 \textmu m detector and its Fourier transform.}
    \label{fig3}
\end{figure*}

We evaluated the performance of the 2 \textmu m T2-HPT as a digital receiver, using bit error rate (BER) measurements (see Method). We coupled the output beam of a 1.55 \textmu m laser diode, which was directly modulated by pseudorandom binary sequences (PRBS), to the integrated grating couplers. Without any external amplifier, we achieved high-quality open eye diagrams at 3.2 Gbps —the limit of our BERT system— as shown in Fig. \ref{fig3}a. We measured sensitivities of -26 dBm and -25 dBm at 2.5 Gbps and 3.2 Gbps respectively, for a $BER<10^9$ (Fig. \ref{fig3}b). More importantly, the high gain and direct driving capability of our device eliminated the need for a high-gain and low-noise amplifier in these experiments and we directly coupled our device to the 50-Ohm load. Since our BERT system was limiting our measured data rate to 3.2 Gbps, we evaluated the actual bandwidth and the resulted estimated bit rate, using a fast optical pulse and a fast oscilloscope (see Methods).
Fig. \ref{fig3}c shows the full width half maximum (FWHM) of the output signal of a 2 \textmu m T2-HPT when receiving an optical pulse with FWHM of $\sim$7 ps. The electrical signal on a 50 Ohms load shows a FWHM of 97 ps and a jitter of less than 9 ps. The corresponding 3dB bandwidth is about 5 GHz, based on the Fourier transform of the output pulse \cite{16_huang201625}. This value is in close agreement with our simulation (see section III of Supplementary). The measured SNR at the output pulse is about 22.5, suggesting a data rate of 10 Gbps at $BER<10^9$ (well beyond the bitrate limit of our BERT system). Even using the measurement-limited data rate of 3.2 Gbps, the 2 \textmu m detector with no external amplifier represents a data rate density of 800 Tbps/mm\textsuperscript{2}, which is significantly higher than that of the best reported amplifier-free APDs ($\sim$100 Tbps/mm\textsuperscript{2} in Ref. \cite{9_virot2014germanium}).
A high data rate density is crucial for the future on-chip optical interconnects \cite{11_miller2009device}, enabling the next-generation computing and sensing chips with massive data bandwidths. 

\section{Discussion}\label{sec3}

\begin{figure*}[t]
    \centering
    \includegraphics[width=12.5cm]{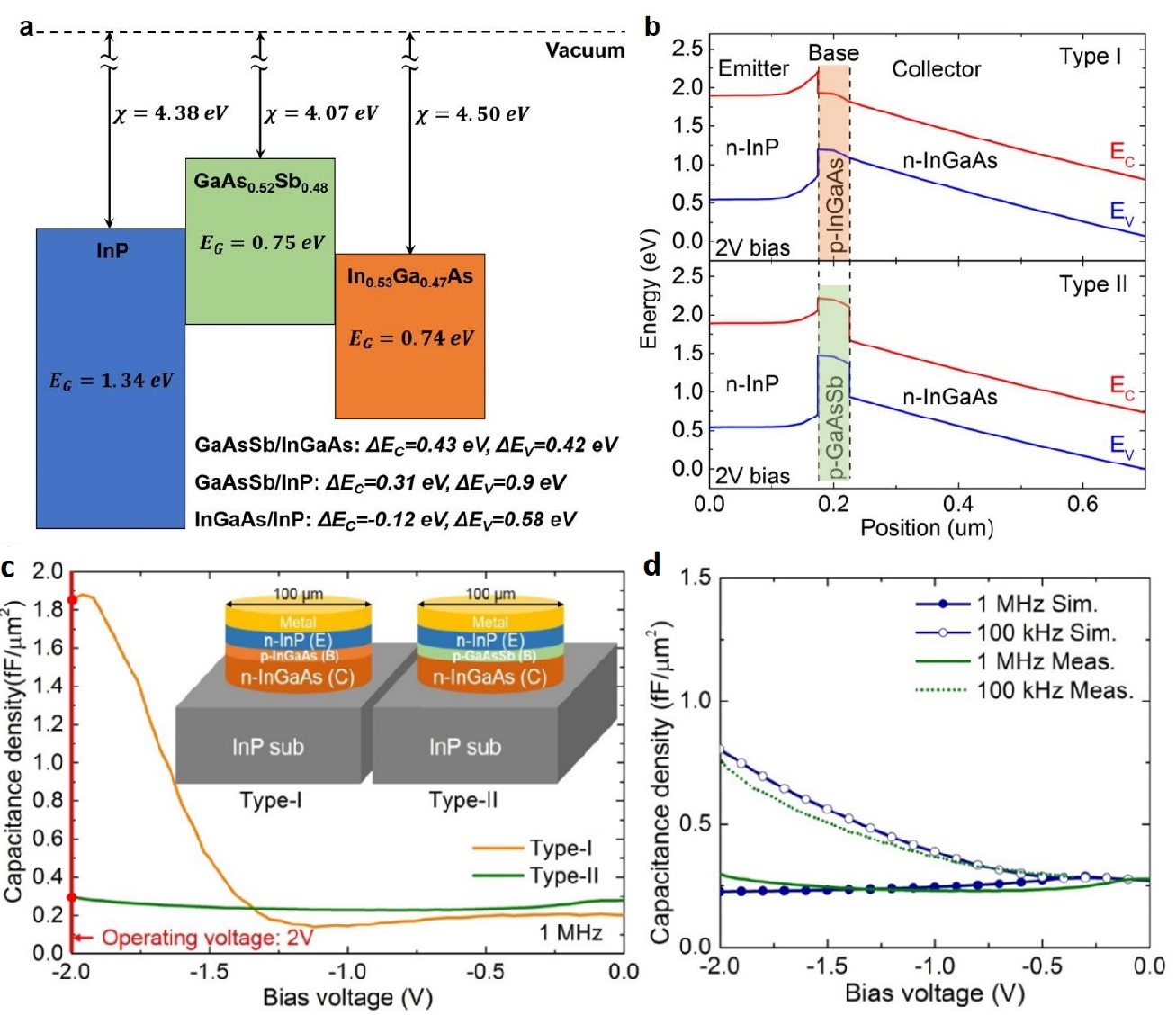}
    \caption{Band structure of type I and type II. \textbf{a}, Schematic diagram of the energy band alignment of InGaAs, GaAsSb, and InP. GaAsSb forms a type-II alignment with InGaAs and InP, while InGaAs/InP forms a type-I alignment. \textbf{b}, Simulated band structures of two N-P-N HPTs based on type-I (top) and type-II (bottom) band alignments. \textbf{c}, Experimentally measured junction capacitance versus bias (C-V) for type-I and type-II HPTs with identical layer thickness and doping levels show that type-II HPT has a substantially lower junction capacitance at the operating bias of -2 Volt. \textbf{d}, Simulated C-V show good agreement with experimental data for low and high frequencies.}
    \label{fig4}
\end{figure*}

It is known that type-II band alignment helps in preventing the onset of the base pushout at high current densities, known as the Kirk effect \cite{bianconi2019engineering}. However, we hypothesized that type-II band alignment can also be used to decrease the device junction capacitance significantly. Reducing junction capacitance of HPTs not only makes the phototransistors faster, but also more sensitive, as we had proposed \cite{39_rezaei2017sensitivity} and recently demonstrated experimentally \cite{41_liu2020highly}.
Fig. \ref{fig4} presents the difference between type- I and type-II band alignment: the discontinuity of the type-II alignment provides a clear advantage from the charge storage point of view, since it prevents excess holes from traveling into the collector and precluding the onset of the Kirk effect. \cite{29_campbell1981small} In addition, a type-II band alignment helps reducing the formation of an electrostatic barrier inside the collector, resulting in a slower increase in capacitance at increased current levels. To evaluate this hypothesis, we measured the capacitance of the type-II HPT and a type-I HPT with equal base doping level, thickness, and diameter. Our experimental results show that type-II HPT exhibits about six times smaller capacitance per unit area than type-I HPT at the bias voltage of 2 V utilized in this work (see Methods). The measured capacitance values and their bias dependencies are in good agreement with numerical simulations for both type-I and type-II devices (see Methods). 

\begin{figure*}[t]
    \centering
    \includegraphics[width=12cm]{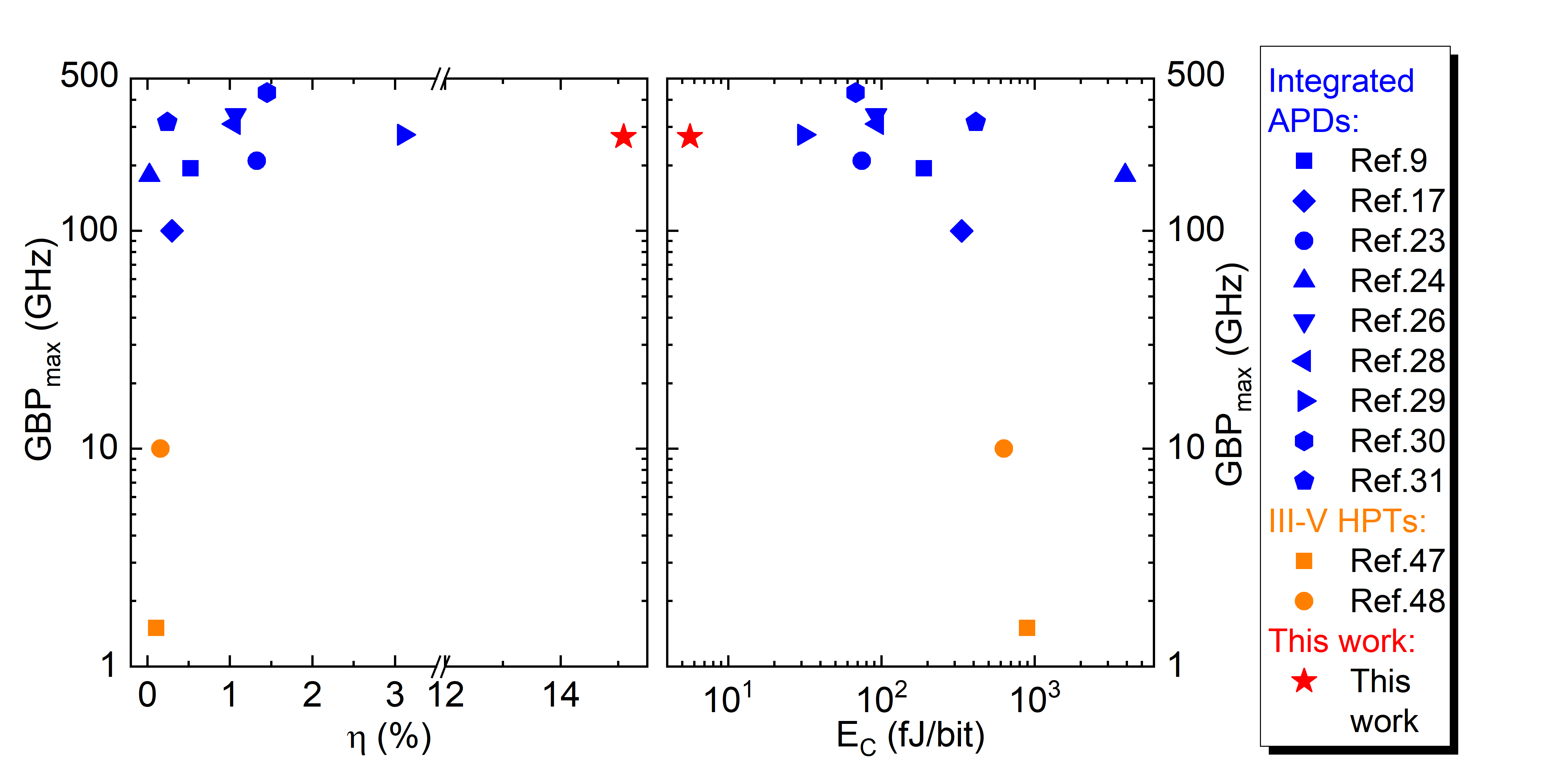}
    \caption{Comparison of this work with the best reported detectors for optical receivers. \textbf{a}, GBPs vs. energy efficiency ($\eta$). Note that we removed a portion of the x-axis, from 2\% to 12\%, to increase the visibility. \textbf{b}, GBPs vs. total consumed energy ($E_C$) of detectors at internal gain of 15 and 2 fF load at optical power of about -30 dBm (see detailed calculation and tabulated data in Supplementary VI).}
    \label{fig5}
\end{figure*}

In addition to the reduced capacitance, our devices are made with special attention to the high current density needed to directly drive loads without an amplifier. Numerical simulations show that compared to our previously reported devices on native substrate, the bonding method and the small dimensions of the devices reported in this work significantly enhance their thermal conductance and reduce their internal temperature (see Supplementary IV for thermal modeling). At an output power of 2 mW (2 V bias voltage and $/approx$1 mA photocurrent), the temperature increase in the transferred detector is more than 4 times lower than that in the as-grown detector. A high temperature can degrade the GBP of device, mainly due to the increased base transit time \cite{42_bailbe1995theory}, as previously demonstrated for type-I and type-II HBT \cite{43_hafizi1993temperature,44_bolognesi2001inp}.
A direct consequence of a fundamentally lower junction capacitance and operating voltage is better energy efficiency. While the presented device is not truly optimized, we would like to compare its energy efficiency with some of the best reported APD and HPT devices. The energy efficiency of an optical receiver can be calculated as $\eta = E_U/(E_C+E_U)$, where $E_C$ is and the energy waste per bit, and $E_U$ is the energy delivered to the load per bit. For approaches based on optical receivers in CMOS, these values can be estimated as (See Supplementary V):

\begin{equation}
    E_U \approx \frac{1}{2} C_L V_L^2
    \label{eq1}
\end{equation}
\begin{equation}
    E_C \approx \frac{ 2 V_b I_d + V_b I_{ph} + 2 V_L I_d + V_L I_{ph}}{2 BR}
    \label{eq2}
\end{equation}

where $C_L$  and $V_L$ are the load capacitance and voltage (around 1 V in modern CMOS), $I_d$, $I_{ph}$ and $V_b$ are the dark current, photocurrent and voltage bias of the photodetector respectively, and $BR$ is the bitrate. From these equations, it is evident that $V_b$ and $I_d$ are the most important factors that determine the energy consumption and efficiency of the receiver. As mentioned above, existing CMOS-compatible APDs require a large $V_b$ to achieve a reasonable avalanche gain and exhibit a large Id due to the large electric field required for avalanche, both of which lead to poor energy efficiencies mostly around 1\%. 
We compare the GBP versus energy efficiency of the best reported results from optical receivers, including III-V HPTs and CMOS-compatible APDs \cite{9_virot2014germanium,17_chen2015high,23_benedikovic202040,24_zeng2019silicon,26_kang2009monolithic,28_thuret1999high,29_campbell1981small,30_fritzsche1981fast,31_chandrasekhar1991demonstration,45_duan2012310,46_huang201625,47_martinez2016high,48_hawkins1997high}. Fig. \ref{fig5}a shows the GBP versus energy efficiency $\eta$ and Fig. \ref{fig5}b shows the GBP versus consumed energy per bit $E_C$ (details in Supplementary VI). Compared to integrated APDs, our integrated T2-HPT exhibits about one order of magnitude higher energy efficiency of $\eta \sim 15\%$, and a significantly lower energy consumption, below $E_C \sim 6$ fJ/bit. Most reports have shown integrated APDs exceeding 10 Gbps bitrates using amplifiers. However, the addition of amplifier circuits has limited the ability to achieve the longstanding goals of high energy efficiency and high data rate densities. Without any amplifier, the maximum achievable bitrate for a given load is set by the photocurrent produced by the device, and hence by the sensitivity level and responsivity of the device. Since the excess noise factor of APDs grows with their internal gain, the practical responsivity of APDs has been limited to below $\sim$30 A/W. Therefore, without an amplifier the achievable bitrate of APDs would be comparable to that of our device (see detailed discussion in Supplementary X, and Table SII). For applications that require a large number of channels per area but moderate bitrates per channel, our device could be used without an amplifier to achieve a very energy efficient solution within an extremely small physical footprint. 

Furthermore, for applications that require high bitrates per channel, the high GBP of the T2-HPT devices still allows use of simpler, more compact, and more efficient amplifiers. As an example, we evaluated the performance of these devices when integrated with a compact 5-transistor CMOS amplifier (See Supplementary VII for the detail of the amplifier). The T2-HPT was modeled using the Ebers-Moll model, which showed a good agreement with the measured performance across different bias values and optical power (see Fig. S9a). Using the compact amplifier, the receiver could achieve a bitrate of 16.7 Gbps (Fig. S9b). When including capacitive loads of 5 fF or 50 fF, the highest open-eye data rates were $\sim$15 Gbps (Fig. S9c) and $\sim$10 Gbps (Fig. S9d) respectively. Crucially, the compact amplifier (Supplementary VII) adds an energy dissipation of $\sim$ 60 fJ/bit, which is smaller than typical amplifiers used in other integrated receivers with a similar CMOS technology node, while achieving $\sim$10 times higher sensitivity and about two times larger bitrate \cite{49_settaluri2015demonstration}.

\section{Conclusion}\label{sec4}

We demonstrated a type-II heterojunction phototransistor (T2-HPT) with exceptional performance characteristics. Arrays of devices with different sizes were integrated on silicon wafers using a CMOS-compatible wafer bonding and an additive waveguide interconnect method. Electrical and optical characterization show that the integrated T2-HPT can achieve a gain-bandwidth product of $\sim$270 GHz at $\sim$10 \textmu W optical power. This device is the first HPT that simultaneously shows a high speed and high sensitivity. When used as an optical receiver, with and without an amplifier, the device showed bitrate and sensitivity values that are similar to the best APDs. However, the T2-HPT showed about 10 times higher energy efficiency and significantly lower energy consumption per bit compared with APDs, due to its extremely low dark current and operating voltage.
Our experimental and simulation results support the hypothesis that the type-II band alignment is potentially the reason for achieving exceptionally high responsivity and low junction capacitance in our device. In parallel, these tiny devices show very large internal gain and current densities required for directly driving large capacitive loads, leading to a massive data transmission density in excess of 800 Tbps/mm\textsuperscript{2} at an attractive energy consumption of $\sim$6 fJ/bit. The unique combination of compactness, low energy consumption per bit, and high energy efficiency makes these new devices a promising choice for high-density optical receivers. We hope that our results encourage the research community to further evaluate type-II phototransistors in new material systems such as van der Waals heterostructures \cite{50_shin2020ultrasensitive}, which present an excellent opportunity for application of this strategy to a broad set of material and wavelengths. 

\section*{Methods}

\subsection*{Fabrication process}

The T2-HPT detectors used in this work is grown by metalorganic chemical vapor deposition (MOCVD) on n-doped InP substrates. Fig. S3 (Supplementary Section II) shows the schematic diagrams of the process flow, which starts with the degreasing of a $\sim$ 1 cm × 1 cm sample and a $\sim$ 1.7 cm × 1.7 cm silicon substrate in organic solvents. Then, $NH_4 OH$ and ($NH_4)_2 S$ treatment are performed on the T2-HPT sample to remove the native oxide and passivate the surface; meanwhile the silicon substrate is dipped in buffered oxide etchant (BOE) for native oxide removal. Both samples are then immediately transferred to the chamber of an electron-beam evaporator, followed by the evaporation of Mo/Au (10/10 nm) bilayer. After an Ar-based plasma treatment is used for surface activation, the two samples are bonded to each other by a wafer-bonding tool (FC-150 by SET) at 200 $\degree$C, and with a force of 35 kg. Then, HCl (37\%) is used to remove the InP substrate and leave behind the epitaxially grown layers. Subsequently, the T2-HPT devices are defined by conventional photolithography and a two-step etching process (dry followed by wet etching). To passivate the sidewalls of the detectors, SiO2 (300 nm) is deposited by Plasma Enhanced Chemical Vapor Deposition (PECVD) at 300 $\degree$C. Keeping the sample inside the PECVD chamber, the hydrogenated amorphous silicon (a-Si:H) waveguide stack is then deposited. It consists of a 400 nm lower cladding layer (RF 40 W, 900 mTorr), a 300 nm core layer (RF 50 W, 300 mTorr) and an 800 nm upper cladding layer (RF 40 W, 900 mTorr). Waveguides with width of 5 \textmu m and grating couplers with period of 670 nm are etched to the lower cladding layer using reactive-ion etching (RIE) and inductively coupled plasma etching (ICP), respectively. Benzocyclobutene (BCB) is then spin-coated, cured and etched back for passivation and planarization. Finally, the Ti/Ni/Au (20/30/100 nm) top and bottom contacts are evaporated after the etching of openings. The temperature of the whole fabrication process never exceeds 300 $\degree$C.

\subsection*{Characterization and measurement}

The electrical and optical DC performance of the fabricated HPTs are characterized by utilizing a 1.55-\textmu m laser diode and a digital multimeter (34410A), connected with a low-noise current preamplifier (SR 570) at room temperature. The light is delivered to the detectors through the waveguide by focusing the beam to the grating coupler. For laser power calibration, the incident optical power through the objective lens was measured using a commercial InGaAs-based p-i-n photodiode module in a dark environment. 
Pulse measurement is conducted using Calmar Optcom Femtosecond pulse laser as the light source emitting an optical pulse with a full width half maximum (FWHM) of 7.31 ps. The output of the device is probed with a GSG probe and directly measured with an Agilent infiniium DCA-J oscilloscope. The FWHM and jitter of the output signal are calculated by the oscilloscope. We assumed that the optical pulse was fast enough to present an impulse input and used Fast Fourier transformation to calculate the frequency response and the device bandwidth. Note that our method underestimates the device bandwidth as it ignores the input pulse width. 
Pseudorandom binary sequence is generated by an Optellent OptoBERT 3200 (Data rate ranges from 155 Mbps to 3.2 Gbps) generator to drive the 1.55-\textmu m laser diode through a Mach-Zehnder amplifier and modulator, the modulated optical signal is then coupled to the integrated T2-HPT. A tunable attenuator is used to change the coupled optical power. The output of the device is probed with a GSG probe and sent to an Agilent infiniium DCA-J oscilloscope and a BERT analyzer to get the eye diagram and Q factor. Note that our BERT system is limited to a maximum data rate of 3.2 Gbps. 
Capacitance-voltage (C-V) characteristics of HPTs with type-I and type-II band alignment are measured by Agilent LCR meter HP 4285. The samples are prepared from the as-grown epitaxial structures on InP substrate, and pillars with diameter of 100 \textmu m are etched and metallized. As shown in the inset of Fig. S4, n-doped InGaAs and InP constitute the collector (C) and emitter (E), respectively. P-doped InGaAs and p-doped GaAsSb constitute the base (B) for type-I and type-II, respectively; both are 50 nm thick and have doping concentration of 5 × 1017 cm-3.

\subsection*{Simulation}

Three-dimensional FDTD simulation was performed using an FDTD tool (Lumerical), which was also used for calculating the transmission of light to the integrated detectors. The device simulation was performed using ATLAS simulation software package. 
Simulations for integrating our T2-HPT detector with a 65nm ASIC amplifier is conducted in Cadence Virtuoso 6.1.8 to evaluate the performance of the detector for applications as an optical receiver (detailed in Supplementary).

\section*{Acknowledgement}

This work was partially supported by ARO award W911NF1810429, NIH award R21EY029516, and the W.M. Keck Foundation Award. The fabrication made use of the NUFAB facility of Northwestern University’s NUANCE Center, and of the Pritzker Nanofabrication Facility part of the Pritzker School of Molecular Engineering at the University of Chicago, which have received support from the Soft and Hybrid Nanotechnology Experimental (SHyNE) Resource (NSF ECCS-1542205); the MRSEC program (NSF DMR-1720139) at the Materials Research Center; the International Institute for Nanotechnology (IIN); the Keck Foundation; and the State of Illinois, through the IIN; and of the Center for Nanoscale Materials of Argonne National Laboratory. Use of the Center for Nanoscale Materials, an Office of Science user facility, was supported by the U.S. Department of Energy, Office of Science, Office of Basic Energy Sciences, under Contract No. DE-AC02-06CH11357.
We would like to thank Moshe Dolejsi (the University of Chicago), and David Czaplewski (Argonne National Laboratory), for their help in optimizing the EBL processes and material etching. We would like to thank Alan Prosser (Fermilab) for his help in the BER measurement and the gigaBERT system.

\section*{Author contributions}

L.L did the designing, processing, measurements and simulations of the integrated T2-HPT detectors. S.B. did the electron beam lithography (EBL) processing, the band energy simulation and the speed modeling. S.W. put together the setup for high-speed measurement. L.L and S.B. wrote the manuscript. N.C. and F.F. did the circuit level simulation. H.M. conceived the idea, guided both experimental and modeling works and revised the manuscript. All authors discussed the results and commented on the manuscript.

\section*{Data availability}

The data that support the findings of this study are available from the corresponding author upon reasonable request.
Competing financial interest
The authors declare no competing financial interest.


\bibliography{sn-bibliography}% common bib file

\end{document}

\title[Article Title]{Supplementary Information: Fast and Efficient Type-II Phototransistors Integrated on Silicon}

\author[1]{\fnm{Lining} \sur{Liu}}
\author[1]{\fnm{Simone} \sur{Bianconi}}
\author[1]{\fnm{Skylar} \sur{Wheaton}}
\author[1]{\fnm{Nathaniel} \sur{Coirier}}
\author[2]{\fnm{Farah} \sur{Fahim}}
\author[1]{\fnm{Hooman} \sur{Mohseni}}\email{hmohseni@northwestern.edu}

\affil[1]{\orgdiv{Bio-Inspired Sensors and Optoelectronics Laboratory}, \orgname{Northwestern University}, \orgaddress{\street{2145 Sheridan Rd}, \city{Evanston}, \postcode{60208}, \state{Illinois}, \country{USA}}}
\affil[2]{\orgdiv{ASIC Development Group, Particle Physics Division}, \orgname{Fermi National Accelerator}, \orgaddress{\city{Batavia}, \postcode{60510}, \state{Illinois}, \country{USA}}}

\maketitle

\tableofcontents
\bigskip
\bigskip
\bigskip

\section{Thin-film transfer procedure and fabrication process}\label{S1}

The epitaxial structure shown in Table \ref{tabS1} was grown on InP substrates. Large-area ($\sim$ 1 cm × 1 cm) epitaxial layers were successfully transferred to silicon substrates by metal-assisted wafer bonding and subsequent wet etching of the InP substrate. Au-Au bonding has been shown to provide a shear bonding strength as high as 20 MPa under a low-temperature process that prevents residual stress responsible for bowing or cracks \cite{S1_32_park2018ingaas,S2_yu2006low,S3_higurashi2009surface}. Mo has a high thermal conductivity (139 W/mK), an excellent contact resistance for III-V semiconductors, and can be used as diffusion barrier to prevent the unfavorable diffusion of Au atoms \cite{S4_baraskar2009ultralow}. Therefore, Mo/Au is chosen as the metal stack to assist the wafer bonding. The mirror-like surface of the transferred film is crack-free and continuous as shown in Fig. 1a in the manuscript. This corroborates the high-quality of the wafer-bonding and chemical InP substrate removal.

\begin{table}[h]
    \centering
    \begin{tabular}{ccc}
        \hline\hline
         Material & Doping & Thickness\\
         \hline
         In\textsubscript{0.53}Ga\textsubscript{0.47}As & N-type, $10^{19} cm^{-3}$ & 100 nm\\
         InP & N-type, $10^{17}$ to $10^{19} cm^{-3}$ graded & 50 nm\\
         InP & N-type, $10^{17} cm^{-3}$ & 200 nm\\
         GaAs\textsubscript{0.52}Sb\textsubscript{0.48} & P-type, $5 x 10^{17} cm^{-3}$ & 50 nm\\
         In\textsubscript{0.53}Ga\textsubscript{0.47}As & N-type, $10^{15} cm^{-3}$ & 1000 nm\\
         In\textsubscript{0.53}Ga\textsubscript{0.47}As & N-type, $10^{15}$ to $10^{18} cm^{-3}$ graded & 50 nm\\
         InP substrate &  & \\
         \hline\hline
    \end{tabular}
    \caption{Epitaxial structure of the HPT wafer grown on InP substrates by low-pressure metal-organic chemical vapor deposition (LP-MOCVD).}
    \label{tabS1}
\end{table}

\begin{figure*}[h]
    \centering
    \includegraphics[width=12.5cm]{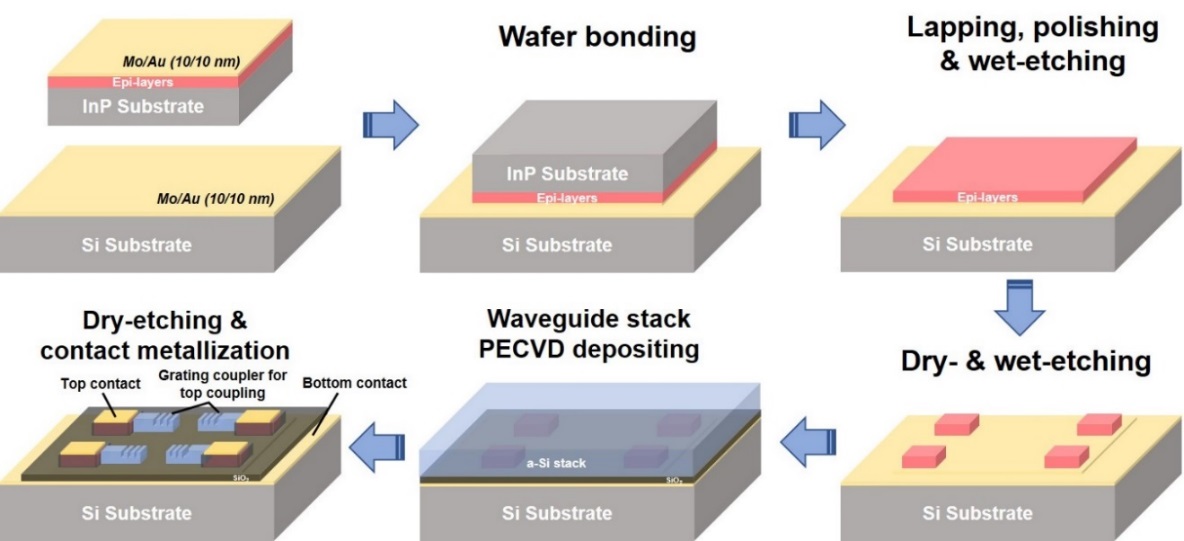}
    \caption{Fabrication process of the integrated T2-HPT detector.}
    \label{figS1}
\end{figure*} 

Table \ref{tabS1} shows the epitaxial structure of the HPT device in this work which uses Sb-based type-II band alignment. The fabrication process of the integrated detectors, waveguides and grating couplers is depicted in Fig. \ref{figS1} and described in the Method section in the manuscript. 
The a-Si:H waveguides and grating couplers consist of a cladding/core/cladding layer stack with a refractive index profile of 2.4/2.8/2.4 at 1.55 \textmu m. a-Si:H is ideal for the optical waveguides and couplers used in this work, due to its high refractive index and low optical loss. It can be easily deposited by plasma-enhanced chemical vapor deposition (PECVD) at low temperatures ($leq$ 300 $\degree$C). Besides, its refractive index can be controlled by adjusting the RF power and chamber pressure during deposition.

\section{Dark current of detectors with different sizes}\label{secS2}
 
\begin{figure*}[h]
    \centering
    \includegraphics[width=10cm]{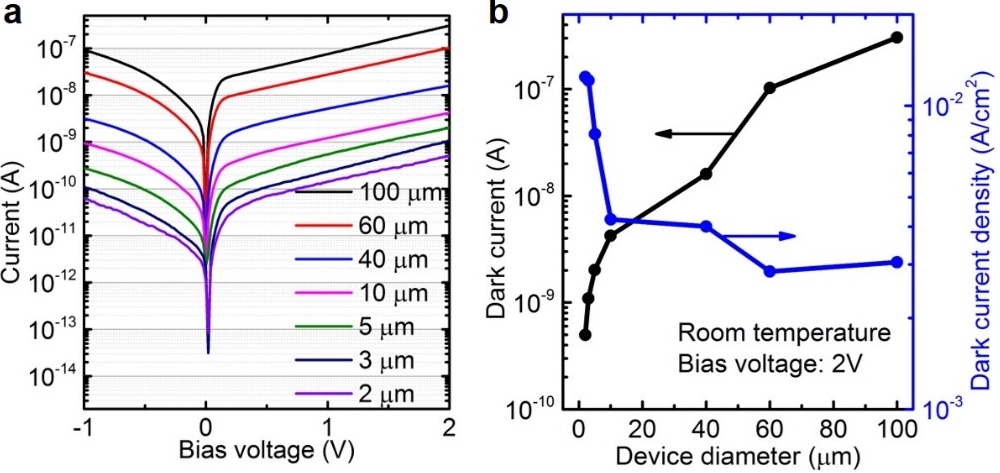}
    \caption{Dark current. \textbf{a}, Dark current versus bias voltage for detectors with different sizes. \textbf{b}, Dark current and dark current density at 2 V bias extracted from \textbf{a}.}
    \label{figS2}
\end{figure*} 

Fig. \ref{figS2}a shows the dark current of detectors with different sizes at room temperature. The detectors with diameter smaller than 10 \textmu m exhibit dark current in nA level at 2 V, which is promising for low energy applications. 
The dark current density (Fig. \ref{figS2}b) increases with decreasing detector size, especially under 10 \textmu m. This is likely caused by the surface effect playing a more dominant role in devices of smaller dimension \cite{S5_dayeh2007transport}.

\section{Model for -3 dB bandwidth of detectors with different sizes}

The -3 dB bandwidth of a device is related to its response time as $BW_{3dB} \approx 0.35/\tau_r$ \cite{S6_orwiler1969oscilloscope}. It has been shown that the rise time of HPTs can be effectively approximated by \cite{S13_7_miller1997physical}:

\begin{equation}
    \tau_r = 2.2 (\tau_{R_L} + r_d C_T + R C_L)
    \label{eqS1}
\end{equation}

where $\tau_{R_L}$ is the recombination lifetime in the base, $r_d$ is the transistor’s dynamic resistance, $C_T$ is the total junction capacitance and the last term represents the RC time constant of the load driven by the detector. By comparing this equation to the measured response time and -3dB bandwidth for devices of different sizes, ranging from 2 \textmu m to 100 \textmu m in diameter, it is possible to separate the contribution of the device RC time constant in the equation above, since it is the only term that depends on the device area. The recombination lifetime is assumed constant for all devices, since the recombination sites are mainly localized at the interface between the base and the emitter layers \cite{S14_11_miller2009device}. Upon shrinking the size of the devices, surface states can provide additional recombination sites, which could explain the slight decrease in response time for the 2 \textmu m device; however, this effect was not noticed in any of devices larger than 2 \textmu m in diameter. The junction capacitance of the devices was approximated by a plate parallel capacitor, as $C = (\varepsilon_0 \varepsilon_r A)⁄w$, where $\varepsilon_0$ and $\varepsilon_r$  are the vacuum permittivity and the semiconductor dielectric constant respectively, and $A$ is the cross sectional area of the devices. The depletion width $w$ of the devices was estimated using a commercial numerical simulation software package (ATLAS from Silvaco International). The dynamic resistance was calculated as \cite{S8_helme2007analytical}:

\begin{equation}
    r_d = \frac{\eta_F k T}{q} \frac{1}{i_D+I_{ph}}
    \label{eqS2}
\end{equation}

where $eta_F$ is the ideality factor, k is Boltzmann’s constant, $T$ is temperature, $q$ is the electron charge, $i_D$ and $i_{ph}$ are the devices internal dark and photo-current, respectively. The ideality factor was calculated based on the optical power level, using the expression proposed by Movassaghi et al. \cite{S9_movassaghi2017analytical} for a class of very similar devices, yielding $\eta_F \approx 2.6$. 

\begin{figure*}[h]
    \centering
    \includegraphics[width=8cm]{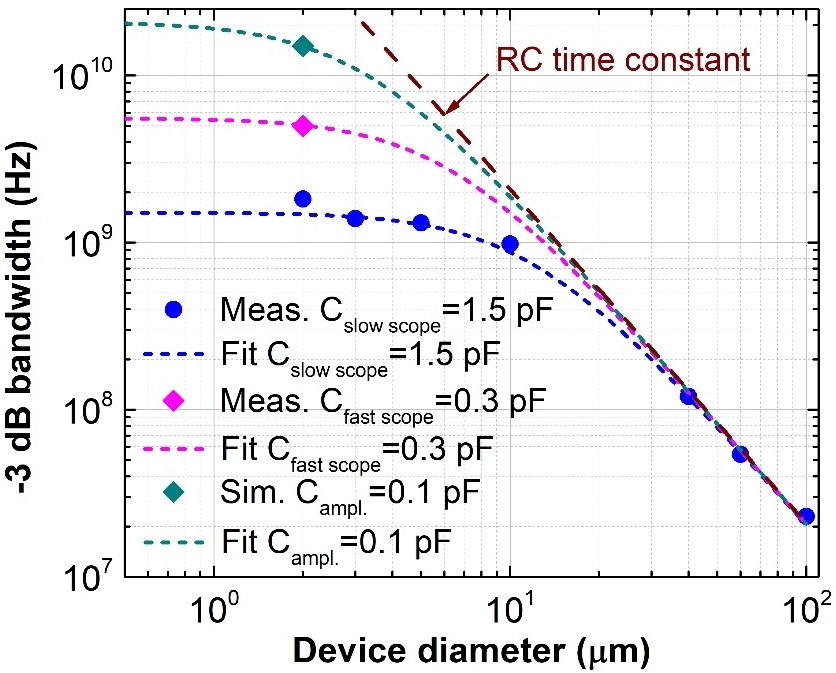}
    \caption{Photoresponse bandwidth model of detectors with different sizes and its fitting to measurement data.}
    \label{figS3}
\end{figure*}

The results of the model are represented by the dashed curves in Fig. \ref{figS3}, for three different values of the load capacitance, corresponding to different measurements and simulations. In this plot, the device RC time constant as a function of its diameter is represented by the brown dashed line. Larger devices are characterized by larger capacitance, and are therefore limited in bandwidth by their RC time constant. Scaling the devices to small sizes allows to decrease the RC contribution due to their response time, and their bandwidth becomes ultimately limited by the recombination lifetime and the capacitance of the load. The smallest devices we measured (2 \textmu m) are limited in bandwidth by the load capacitance of the measurement instrumentation. Our experimental results show reducing the capacitance of the oscilloscope results in an improvement in bandwidth, as shown in Fig. \ref{figS3}. In addition, integrating the devices with a low-capacitance ASIC amplifier could further increase their -3dB bandwidth, as shown by our detailed simulation using Cadence Virtuoso, and represented by the green diamond in Fig. \ref{figS3}. Note that the high internal gain of the T2-HPT allows a very compact and simple amplifier design, with very low power consumption and small footprint. The details of the simulation and ASIC design are given in the following section.

\section{Thermal modeling of the integrated and as-grown detector}
 
Table \ref{tabS2} compares the GBP and speed of devices we previously reported to those achieved in this work, which are significantly superior \cite{S10_bianconi2019engineering,S11_fathipour2017detector}. While in this work we have fabricated devices smaller than ever before, we note that size scaling is not the only factor improving the performance of the detector. For example, Device \#2 in Table \ref{tabS2} has the same epitaxial structure as the ones used in this work, however, when comparing detectors of identical size of 30 \textmu m, the GBP in this work is about 6 time higher. The devices in our previous reports present significant differences compared with the ones presented in this work. Crucially, in our previous work, the detectors were fabricated on the InP native substrate (as-grown) and back-illuminated, while in this work the epitaxial layer is transferred to a silicon substrate and subsequently fabricated on silicon, and side-illuminated through integrated a-Si waveguides.

\begin{table}
    \centering
    \begin{tabular}{m{3cm} m{2.5cm} m{2.7cm} m{2.5cm} }
        \hline\hline
        Device  & GBP max (GHz) & Minimum measured rise time (ns) & Integrated on Si\\
        \hline
        Previous Device\#1 \cite{S10_bianconi2019engineering} & 1.2 & 32.5 & No\\
        Previous Device\#2 \cite{S10_bianconi2019engineering} & 4.9 & 20 & No\\
        Previous Device\#3\cite{S10_bianconi2019engineering} & 0.9 & 395 & No\\
        Previous Device\#4 \cite{S11_fathipour2017detector} & 6 & 5 & No\\
        This work & 270 & 0.04 & Yes\\
         \hline\hline
    \end{tabular}
    \caption{Comparison of this work with our previous published works based on HPT}
    \label{tabS2}
\end{table}

\begin{figure*}[t]
    \centering
    \includegraphics[width=10cm]{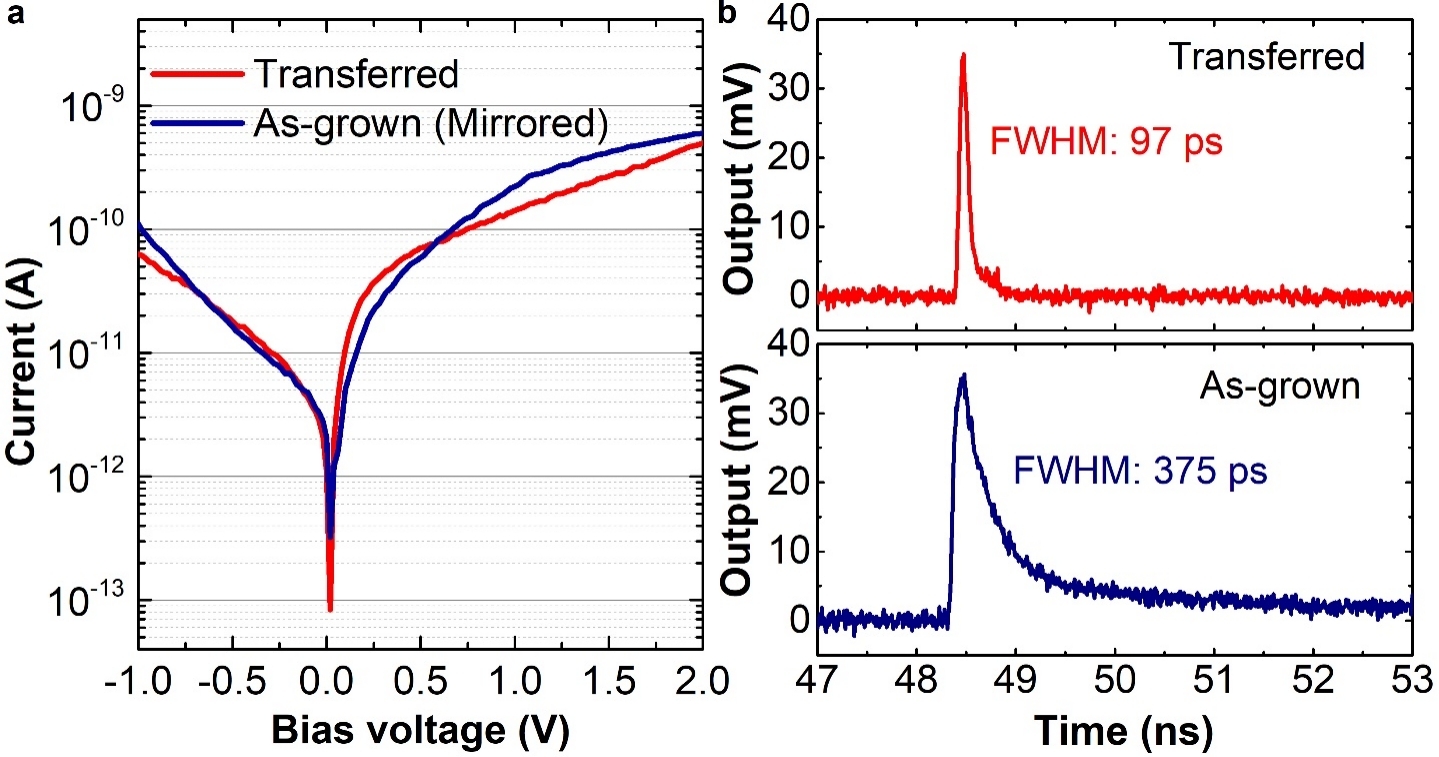}
    \caption{Comparison of transferred and as-grown HPT detectors. Comparisons of \textbf{a}, dark current and \textbf{b}, output pulses from the transferred and as-grown T2-HPTs. Note that after bonded to Si substrate, the transferred detector is upside down with respect to the as-grown detector, so its bias voltage should be reversed with each other. For better comparison, the dark current curve of the as-grown detector is mirrored in panel \textbf{a}.}
    \label{figS4}
    \vspace{-15pt}
\end{figure*}

\begin{figure*}[t]
    \centering
    \includegraphics[width=8cm]{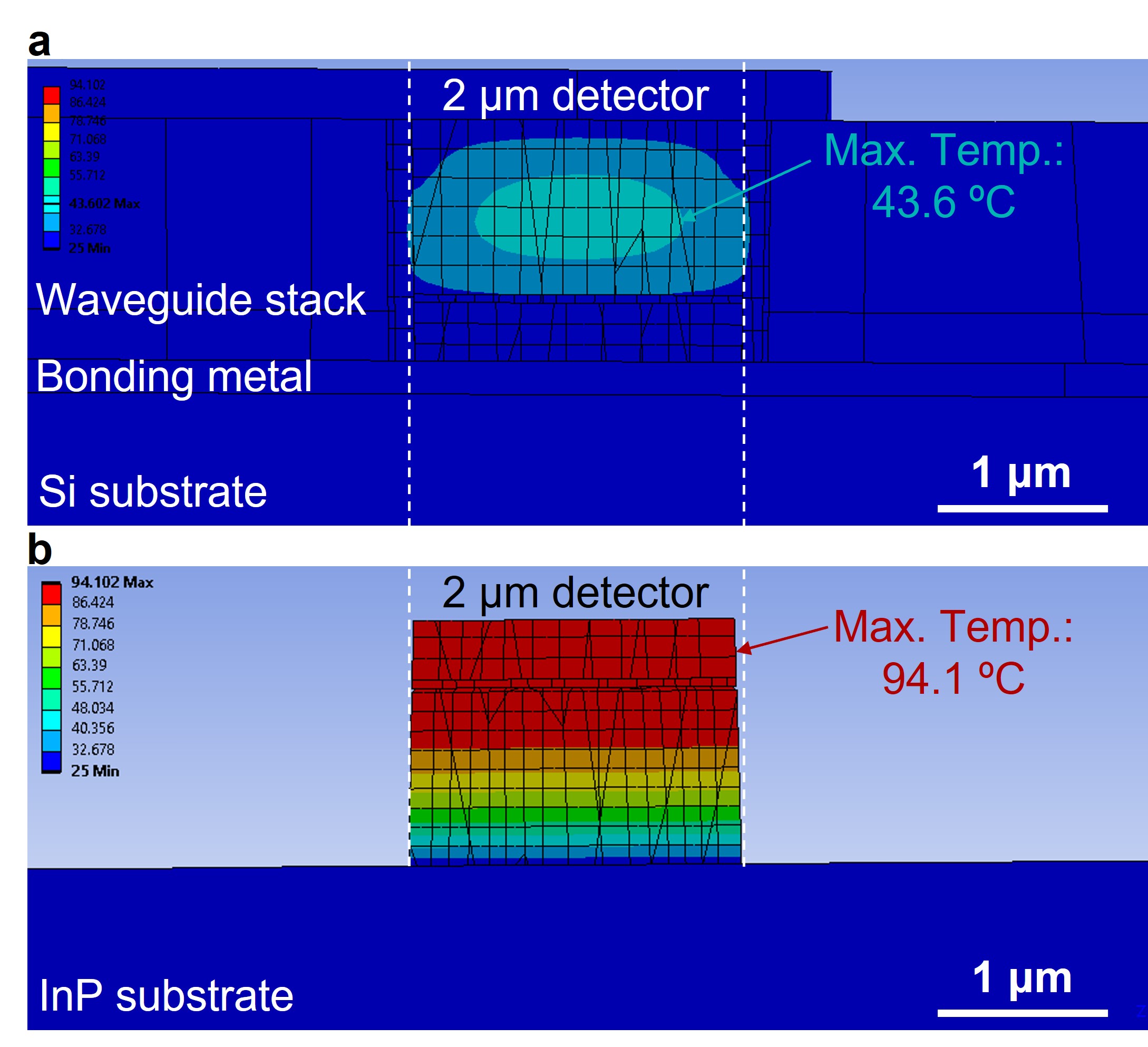}
    \caption{Comparison of thermal conductance between the as-grown and integrated detectors. Thermal simulation of the \textbf{a}, transferred and \textbf{b}, as-grown detectors.}
    \label{figS5}
\end{figure*}

To conclusively compare the integrated and as-grown detectors, we made a sample with 2 \textmu m devices on native InP substrate (as-grown), using the exact same epitaxial material. As expected, the DC performance of the two sets of devices are almost identical. Notably, the dark current of the transferred and as-grown detectors are very similar (Fig. \ref{figS4}a), indicating that transferring the detector to Si substrate has no significant effect on the dark current of the detector. In order to rule-out any possible differences or limitations in the measurement setup, we conducted a femtosecond laser pulse measurements using the same measurement setup both for the transferred and as-grown detectors. When comparing the speed of both devices, we observed that the transferred detector is $\sim$ 4 times faster than the as-grown detector at the same output amplitude, as shown in Fig. \ref{figS4}b. Therefore, with exactly the same size and measurement conditions, the transferred detectors conclusively exhibit higher speed the as-grown detectors. The exact physics behind the marked difference is not clear at this time and will be a subject of our future work. However, we hypothesize that it is related to the significantly higher thermal and electrical conductivity of the back metal contact when devices are transferred to Si substrate, and the lower diffusion length needed when the device is side illuminated rather than back-side illuminated.

In support of this hypothesis, we performed a detailed 3D thermal simulation using ANSYS thermal modeling (FEM) tool to compare the as-grown and integrated detectors. The power generated from the photo current is assumed to be 2 mW, since the photocurrent in Fig. 2d in the manuscript is in mA level and the operating bias voltage is 2 V. Assuming the temperature in the surrounding areas is 25 $\degree$C, the temperature increase during detection in the transferred detector ($\Delta T = 18.6 \degree C$) is significantly lower than that of the as-grown detector ($\Delta T = 69.1 \degree C$), as shown in Fig. \ref{figS5}.

\section{Energy consumption in optical receiver system using photodetectors}
 
\begin{figure*}[h]
    \centering
    \includegraphics[width=10cm]{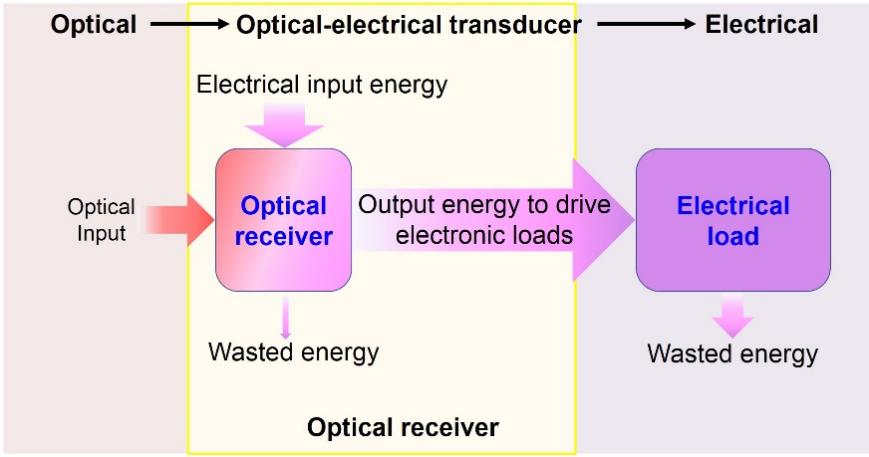}
    \caption{Energy flow of the system for optical interconnects. The optical receiver transduces the input optical signal to an electrical signal for on-chip data transmission. The input energy of the system consists of an optical signal and an electrical bias, part of which is wasted by the dark current of the optical receiver and the rest of which is transduced to the electrical load. In the electrical load stage, part of the energy is utilized while the other part is wasted. Calculation of these energies is discussed in the text.}
    \label{figS6}
\end{figure*}

To better understand the sources of energy waste in an optical receiver, let us first examine the energy flow. In an optical receiver, data is transduced from the optical domain to the electrical domain (Fig. \ref{figS6}). For most practical applications, the energy of the signal must also be boosted significantly since the input optical energy is typically less than the electrical energy required to drive the capacitive load of the connected electronics. This is commonly achieved using p-i-n or avalanche photodiodes (APD) combined with an electrical amplifier.
 
\begin{figure*}[t]
    \centering
    \includegraphics[width=6cm]{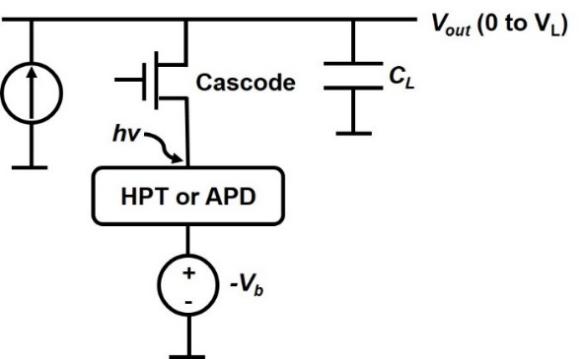}
    \caption{Equivalent circuit for an optical receiver based on a photodetector.}
    \label{figS7}
    \vspace{-15pt}
\end{figure*}

Here we consider the simplest general equivalent circuit that could be used for both HPT and APD detectors to create an optical receiver system. Since the detector always needs a bias Vb, in order to keep the lower side of the load at ground, a negative voltage bias is applied (-$V_b$). A cascode transistor is the simplest circuit to stabilize the detector bias around Vb, which is crucial to maintaining the gain and speed of HPTs and APDs while transferring its current to the load. Here, we assumed an ideal transistor and current source, although these assumption have little effect on the general conclusions that can be made from this model (Fig. \ref{figS7}).

\begin{figure*}[t]
    \centering
    \includegraphics[width=12.5cm]{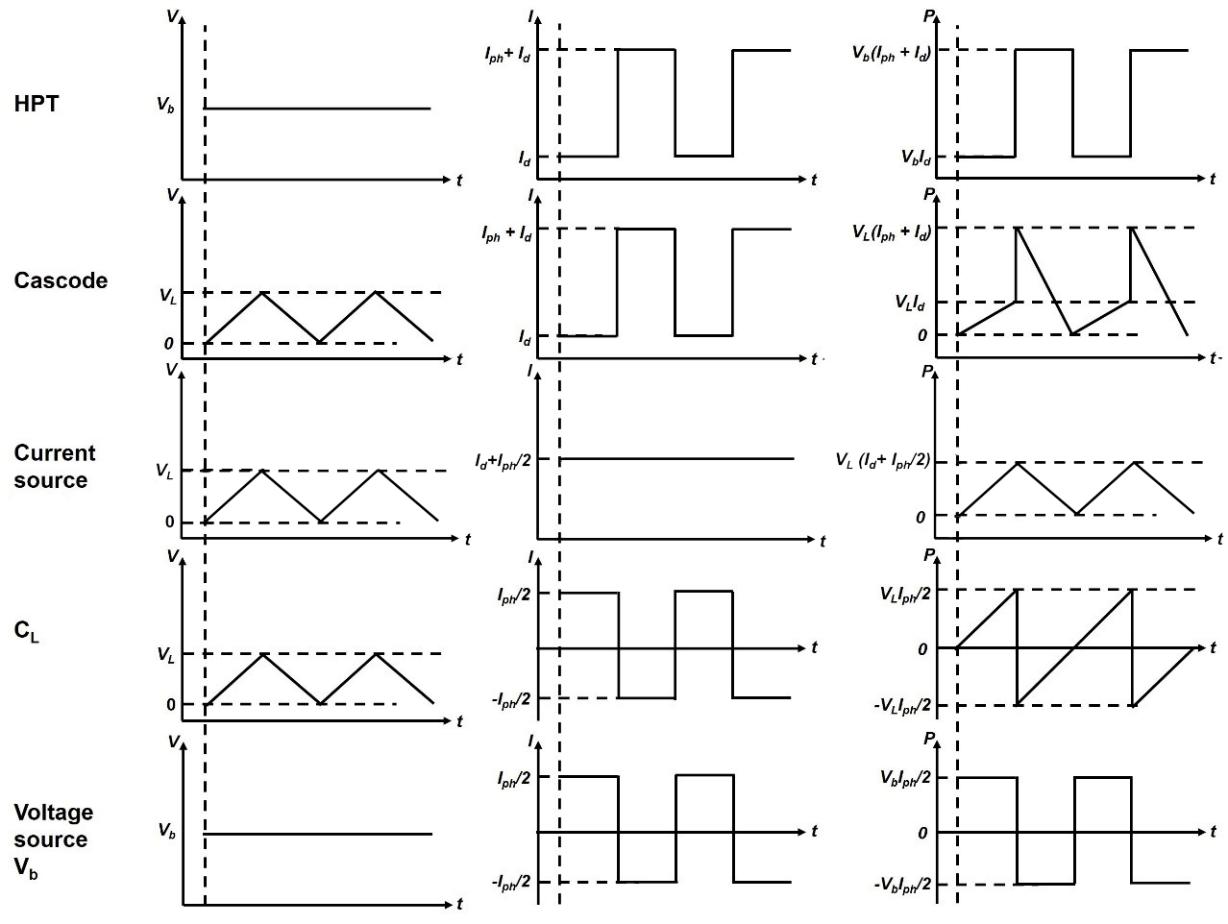}
    \caption{Current, voltage and power consumption of each element in the circuit.}
    \label{figS8}
\end{figure*}

The energy dissipation and efficiency can be calculated from the current and voltage of each element in this circuit. Here we used the conventional approach of assuming an equal probability for the “0” and “1” bits (alternative modulation schemes with unbalanced probabilities result in a similar overall energy conclusion). The first and second column in Fig. \ref{figS8} show the current and voltage of each element respectively. The instantaneous power dissipated by each element (third column) can be obtained from the product of its current and voltage. Therefore, the energy dissipated per cycle for each element (denoted as $E_{detector}$, $E_{cascode}$, $E_{cs}$, $E_{load}$, and $E_{vs}$ for the detector, cascode transistor, current source, load and voltage source, respectively) equals to the integration of its instantaneous power over time (Eq. \ref{eqS3} to \ref{eqS6}).

\begin{equation}
    E_{detector} = \frac{1}{2} \frac{V_b I_{ph} + 2 V_b I_d}{BR} 
    \label{eqS3}
\end{equation}

\begin{equation}
    E_{cascode} = \frac{1}{2} \frac{V_L I_{ph}/2 + V_L I_d}{BR}
    \label{eqS4}
\end{equation}

\begin{equation}
    E_{cs} = \frac{1}{2} \frac{V_L I_{ph}/2 + V_L I_d}{BR}
    \label{eqS5}
\end{equation}

\begin{equation}
    E_{load} = E_{vs} = 0
    \label{eqS6}
\end{equation}

And since the energy utilized by the load $E_U$ is:

\begin{equation}
    E_U  \sim  \frac{1}{2} C_L V_L^2  
    \label{eqS7}
\end{equation}

The receiver energy consumption is:

\begin{equation}
    E_C = E_{detector} + E_{cascode} + E_{cs} = \frac{2 V_b I_d + V_b I_{ph} + 2 V_L I_d + V_L I_{ph}}{2BR}
    \label{eqS8}
\end{equation}

Therefore, the energy efficiency is:

\begin{equation}
    \eta \sim  \frac{E_U}{E_C+E_U} = \frac{\frac{1}{2} C_L V_L^2}{\frac{2 V_b I_d + V_b I_{ph} + 2 V_L I_d + V_L I_{ph}}{2BR} + \frac{1}{2} C_L V_L^2} 
    \label{eqS9}
\end{equation}

\section{Comparison of the best reported on-chip optical receivers}
 
First we consider the case of photodetectors without internal gain (e.g. p-i-n photodiode), which require external amplifiers. Table \ref{tabS3} compares one of the best reported optical receiver based on p-i-n photodiode integrated with a custom-designed CMOS transimpedance amplifier (TIA) \cite{S12_15_gasser2023highly} with the optical receiver presented in this work, operating without an external amplifier. The latter shows higher bandwidth, much smaller area and one order of magnitude lower power consumption, at a similar sensitivity level.

\begin{table}
    \centering
    \begin{tabular}{m{4cm} m{4cm} m{4cm}}
        \hline\hline
         & State-of-the-art PIN + TIA & Type-II HPT in this work\\
         \hline
         Area & 410 \textmu m × 410 \textmu m & 2 \textmu m in diameter\\
        Power consumption & 31 mW (TIA) & 2 mW (no amplifier)\\
        Data rate with high quality eye diagram with lowest optical power & 2 Gbps with -27.45 dBm & 3.2 Gbps with -27 dBm\\
        3 dB bandwidth & 1.4 GHz & 5 GHz\\
        Waveguide-coupled & No & Yes\\
         \hline\hline
    \end{tabular}
    \caption{Comparison between one of the best reported p-i-n optical receiver with external amplifier and the receiver presented in this work.}
    \label{tabS3}
\end{table}

\begin{table}
    \centering
    \begin{tabular}{b{0.4cm} b{1.9cm} b{0.9cm} b{0.7cm} b{0.7cm} b{1cm} b{1cm} b{0.7cm} b{0.7cm}  b{0.5cm}}
        \hline\hline\\ [0.1ex] 
        Ref. & Type & $GBP_{max}$ (GHz) & R (A/W) & $V_b$ (V) & $I_d$ (A) & $I_{ph}$ (A) & $BR_{max}$ (Gbps) & $E_C$ (fJ/bit) & $\eta$ (\%)\\ [0.3ex] 
        \hline\\ [0.1ex] 
        5 & Ge APD & 193 & 6 & 6.3 & $3.6 \; 10^{-5}$ & $6.0 \; 10^{-6}$ & 1.5 & 190 & 0.52\\ [0.3ex] 
        17 & Ge/Si APD & 100 & 9 & 6.2 & $1.0 \; 10^{-4}$ & $9.0×10^{-6}$ & 2.3 & 334 & 0.30\\ [0.3ex]
        23 & Ge APD & 210 & 7.35 & 9 & $1.0 \; 10^{-5}$ & $7.4 \; 10^{-6}$ & 1.9 & 74.4 & 1.33\\ [0.3ex]
        24 & Ge/Si APD & 180 & 7.2 & 6 & $1.0 \; 10^{-3}$ & $7.2 \; 10^{-6}$ & 1.8 & 3900 & 0.03\\ [0.3ex]
        26 & Ge/Si APD & 340 & 8.25 & 24 & $3.5 \; 10^{-6}$ & $8.2 \; 10^{-6}$ & 2.1 & 92.4 & 1.07\\ [0.3ex]
        28 & Ge/Si APD & 310 & 4.5 & 26.5 & $1.6 \; 10^{-6}$ & $4.5 \; 10^{-6}$ & 1.1 & 94.1 & 1.05\\ [0.3ex]
        29 & Ge/Si APD & 276 & 15.75 & 9.8 & $3.5  \; 10^{-6}$ & $1.6 \; 10^{-5}$ & 3.9 & 31.2 & 3.11\\ [0.3ex]
        30 & Ge/Si APD & 432 & 12 & 28.9 & $8.1 \; 10^{-7}$ & $1.2 \; 10^{-5}$ & 3.0 & 67.9 & 1.45\\ [0.3ex]
        31 & InGaAs APD & 315 & 8.55 & 21.5 & $3.5 \; 10^{-5}$ & $8.6  \; 10^{-6}$ & 2.1 & 413 & 0.24\\ [0.3ex]
        45 & InP/InGaAs HPT & 62 & NA & 1.6 & $1.0 \; 10^{-7}$ & NA & NA & NA & NA\\ [0.3ex]
        46 & InP/InGaAs HPT & 1.7 & NA & NA & NA & NA & NA & NA & NA\\ [0.3ex]
        47 & InP/InGaAsP HPT & 1.5 & 12 & 7 & $5.0  \; 10^{-5}$ & $1.2 \; 10^{-5}$ & 0.5 & 896 & 0.11\\ [0.3ex]
        48 & InP/InGaAs HPT & 10 & 12 & 2 & $1.5 \; 10^{-5}$ & $1.2 \; 10^{-5}$ & 0.1 & 630 & 0.16\\ [0.3ex]
        This work & InGaAs/GaAsSb HPT & 500 & 12 & 2 & $5.0 \; 10^{-10}$ & $1.2 \; 10^{-5}$ & 3.0 & 5.63 & 15.09\\ [0.3ex]
        \hline\hline
    \end{tabular}
    \caption{Characteristics of the best reported photodetectors and this work, assuming all the detectors work at internal gain of 15 with optical power of -30 dBm and load capacitance of 2 fF.}
    \label{tabS4}
\end{table}

Second, we consider the comparison to other photodetectors with internal gain (e.g. APD and HPT): GBP and other characteristics of the best reported optical receivers, needed for the energy calculation discussed in Section V, are listed in Table \ref{tabS4} (the first column in Table \ref{tabS4} corresponds to the reference numbers in the manuscript). Due to the high excess noise factor, most APDs are not able to operate at high gain, and commonly reported results in the literature are limited to a gain of $\sim$15. Therefore, we here use the reported currents and voltages around a gain of 15, when possible, or the maximum gain in order to be able to objectively compare a large number of the best reported devices. In addition, in this comparison we also evaluate our devices in the low-gain regime (i.e. around a gain of 15), in order to be able to compare them to the reported devices. We assumed an optical power of -30 dBm and a capacitive load of 2 fF, which represents a typical power and load (note that CMOS interconnect line capacitance can be approximated as $\sim$ 0.2 fF/\textmu m, almost independent of the technology node and the typical input capacitance of most high-speed circuits is about 1 fF S13,S14). As shown in Section V, the current reaching the load is $I_{ph}/2$, and hence:

\begin{equation}
    I_{ph}/2 = C_L \frac{dV}{dt} 
    \label{eqS10}
\end{equation}

Therefore, the highest current-limited bitrate for each device can be approximated by:

\begin{equation}
    BR_{(max} \sim I_{ph}/( 2 C_L V_L )  
    \label{eqS11}
\end{equation}

The value of $BR_{max}$, shown in the last column in Table \ref{tabS4} is the lower value of this calculation (current limit) and the measured bitrate (device’s bandwidth limit). Looking at the table, it is evident that the highest bitrates of integrated APDs are similar to the integrated T2-HPT when directly driving a capacitive load. When using an external amplifier, the load capacitance can be reduced and hence the bitrate increased. In the next section, we show that the bitrate of our integrated T2-HPT can be boosted to 15 Gbps with a simple amplifier, which is comparable to the bitrate of APDs with integrated amplifiers, but still with a better energy efficiency and data rate density.

\section{Simulation of T2-HPT integrated with an ASIC amplifier}

\begin{figure*}[h]
    \centering
    \includegraphics[width=10cm]{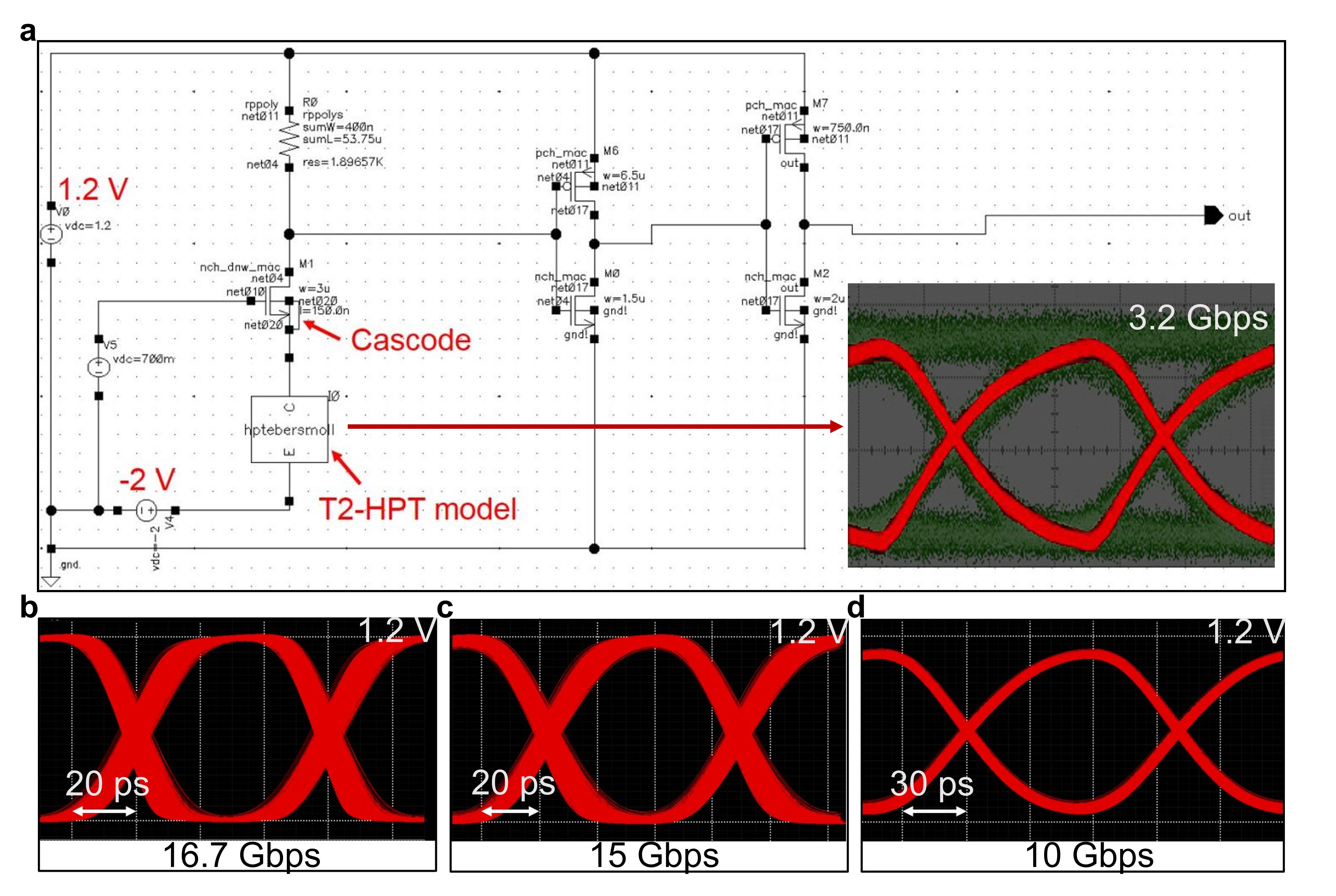}
    \caption{Amplifier with low power consumption design for applications require high bitrate/channel. \textbf{a}, Amplifier circuit used in conjunction with the T2-HPT. The inset shows that the simulated eye diagram at 3.2 Gbps has good agreement with the measured eye diagram shown in Fig. 3a in the manuscript (the voltage swing of 20 mV matches the measured value). \textbf{b, c, d,} Simulated eye diagrams of the output of a simple amplifier integrated with the HPT detector without load (\textbf{b}), and with 5 fF (\textbf{c}) and 50 fF (\textbf{d}) load added between the output and ground, achieving open-eye at bitrates of 16.7 Gbps, 15 Gbps and 10 Gbps respectively.}
    \label{figS9}
\end{figure*}
 
The optical receiver performance of our HPT detector integrated with a 65nm ASIC was simulated using Cadence Virtuoso 6.1.8. In this circuit (Fig. \ref{figS9}a), a forward-active Ebers-Moll BJT model is used for the T2-HPT detector. This model shows very good agreement with our experimental results across different bias values and optical powers. A cascode transistor is then used to pin the HPT’s collector voltage near 0 V, in order to ensure proper device biasing of around 2 V. This cascode acts in conjunction with the resistor and the first inverter stage to form a simple voltage divider/ transimpedance amplifier. The second inverter stage is used to condition the signal such that this circuit can drive a larger load, in addition to adjusting the eye crossing point of the signal much closer to 50\%. The CMOS portion of the circuit is powered by a standard 1.2 V supply, which shares a ground with the -2 V supply used by the detector. By adjusting the Ebers-Moll model’s base-collector and base-emitter capacitance as well as the emitter resistance, eye diagrams that are almost identical to the lab measurement results at 3.2 Gbps were obtained (inset of Fig. \ref{figS9}a). Simulation results show that the highest data rate from the output of this circuit is 16.7 Gbps (Fig. \ref{figS9}b). We simulated the performance of the receiver for a moderate capacitive load of 5 fF and a large capacitive load of 50 fF, and results show data rates of 15 Gbps and 10 Gbps respectively (Fig. \ref{figS9}c and Fig. \ref{figS9}d).

\bibliography{sn-bibliography}